# Temporal Renormalization and the Critical-like Behavior in Supercooled Liquids

D. M. Zhang[a,b], D. Y. Sun[c§] and X. G. Gong[a§]

[a]Key Laboratory for Computational Physical Sciences (MOE), Institute of Computational Physics, Fudan University, 200433 Shanghai, China

[b]Department of Physics, College of Physics and Information Engineering, Fuzhou University, 350108 Fuzhou, China

[c]School of Physics and Electronic Science, East China Normal University, 200241 Shanghai, China

**Abstract**

Inspired by the Kadanoff transformation in the standard renormalization group theory, we propose a temporal renormalization scheme. A Boltzmann factor that explicitly depends on the renormalized timescale is constructed, permitting thermodynamic quantities to be evaluated self-consistently across different timescales. By applying the scheme to the long-time dynamics of supercooled liquids, we uncover critical-like behaviors of supercooled liquid with three characteristic renormalization timescales: At the first timescale $s_\alpha$, the system appears to be "thermodynamically frozen", i.e., the energy fluctuation becomes temperature-independent throughout the supercooled regime. At the second timescale $s_\beta$, the third-order moment of energy distribution reaches a maximum, and $s_\beta$ is nearly temperature-independent. At the third timescale $s_\gamma$, the third-order moment of energy distribution passes through a minimum, and $s_\gamma$ diverges as a power law $s_\gamma \sim |T - T_c|^{-\gamma}$. The scaling relations may reveal an intrinsic behavior in supercooled liquids, highlighting their unique feature. The current findings also demonstrate that temporal renormalization provides a powerful lens for investigating the timescale-specific dynamics.

§Corresponding Authors: xggong@fudan.edu.cn; dysun@phy.ecnu.edu.cn

**Introduction:** In condensed matters, a wealth of fascinating phenomena is inseparable from dynamical or relaxation processes occurring at various timescales [1–9]. The timescale-specific dynamics could also be decisive in some processes or phenomena, which are particularly common and critical in non-equilibrium and complex systems [10–17]. Standard statistical mechanics, however, encodes no explicit time variable in the partition function [18]. As a result, neither the timescales themselves nor their influence on thermodynamic quantities can be read directly from the partition function. Therefore, constructing a formalism that makes this link explicit remains a central challenge [19–21].

For most time-dependent problems, the central obstacle is the accurate treatment of long-timescale processes [22–27]. The long-time dynamical events usually exhibit relatively weak signals, so their signatures are easily overwhelmed by rapid thermal fluctuations and fast processes. Supercooled liquids furnish a paradigmatic example: their dynamical processes span across many orders in time. Long-time correlations are widely held to dictate the thermodynamic signature and, very plausibly, the glass transition itself [28,29]. Yet, in the absence of efficient theoretical methods that map timescale-specific dynamics to thermodynamics, the true nature of the supercooled liquid and glass remains an open question.

Disentangling thermodynamic behavior on timescales is a non-trivial task, whose principal obstacles are (i) the reliable isolation of various timescale processes or relaxations [26,30], and (ii) the faithful projection of such timescale-specific dynamics onto thermodynamic properties [31–33]. At present, to our best knowledge, satisfactory solutions to either problem are unavailable. We note that in traditional renormalization methods, the Kadanoff transformation can highlight thermodynamic behavior at large spatial scales [34,35]. By extending the Kadanoff transformation to the temporal axis, we could obtain a temporal renormalization scheme capable of tracking thermodynamic behavior across different timescales. However, translating this spatial idea to the temporal axis is anything but straightforward. First, the partition function, which does not explicitly contain time variable, offers no natural foothold for a temporal renormalization. Second, the renormalization must leave the long-time dynamics intact; in other words, any temporal renormalization has to preserve, without distortion, the long-time dynamical imprint on thermodynamic quantities.

This work constructs a temporal renormalization formalism. Within this framework, we are able to quantify the long-time dynamical processes of supercooled liquids, and discovery three characteristic renormalization timescales, which should be the intrinsic properties of long-time dynamics in the supercooled liquid.

**Theoretical Framework:** Consider a system characterized by a time series $\{E(t)\}$ representing sequential instantaneous energy recorded throughout its temporal evolution at the temperature of $T$. This energy "trajectory" can be obtained from molecular dynamics (MD) simulations, Monte Carlo simulations, or experimental observations. The occurrence probability $P(E(t))$ of microstates associated with the energy $E(t)$ at time $t$ is

$$P(E(t)) \propto e^{-\beta E(t)}, \tag{1}$$

where $\beta = 1/k_B T$. Considering a temporal transformation $\mathcal{F}$ to the original energy sequence $\{E(t)\}$, which yields a new set of time series:

$$E_s(t') = \mathcal{F}[E(t), E(t+1), \dots, E(t+s)]. \tag{2}$$

Here, $s$ represents the renormalization timescale. If transformation $\mathcal{F}$ can effectively eliminate dynamical processes with timescales shorter than $s$, the new time series $\{E_s(t)\}$ enables the investigation of long-time dynamical behavior. For formal simplicity in subsequent derivations, we adopt a discretized representation of $\{E(t)\}$.

Inspired by the Kadanoff transformation, we propose $\mathcal{F}$ as the following transformation:

$$E_s(i) = \frac{1}{s} \sum_{t=is+1}^{is+s} E(t). \tag{3}$$

Hereafter, this transformation is referred to as $K_s$. This transformation makes the time series into renormalization blocks of length $s$ and assigns the energy $E_s(i)$ to the $i$-th block. While $K_s$ may not be the only admissible realization of $\mathcal{F}$, any transformation preserving long-time dynamics could be viable. However, our analysis demonstrates that $K_s$ provides the most operationally tractable and theoretically minimal formulation. By means of the $K_s$ transformation, we can isolate the dynamical processes whose characteristic timescales exceed $s$. Now the probability of a block microstates composed of $s$ consecutive states associated to the short-time averaged energy $E_s$ can be written as:

$$P(E_s(i)) = \sum_{t=is+1}^{is+s} e^{-\beta E(t)}. \tag{4}$$

Eqs. (3) and (4) do not introduce new physics, they merely redefine the microstates. In this work, our goal is to write $P(E_s(i)) \propto e^{-\beta E_s(i)}$, thus the thermodynamic relations can be directly applied to the energy sequence $\{E_s(i)\}$. Fortunately, under reasonable approximations (details of the mathematical derivation can be found in the Supplementary Materials), we can obtain,

$$P(E_s(i)) = \Theta(s,\beta) e^{-\beta E_s(i)} \propto e^{-\beta E_s(i)}, \tag{5}$$

where $\Theta(s,\beta)$ includes the contribution from short-time (less than $s$) dynamical processes, and is approximately constant and independent of $E_s$. Now $P(E_s(i))$ has the form of the standard Boltzmann factor, from which the thermodynamic quantities of interest can be directly calculated.

Each renormalization operation removes the correlation $\Theta(s,\beta)$ within the $s$ consecutive states, then the contribution from short-time dynamical processes is relegated to the discarded $\Theta(s,\beta)$. In practical applications of our method, the explicit form of $\Theta(s,\beta)$ is not required. Our purpose here is solely to demonstrate that $\Theta(s,\beta)$ is approximately independent of $E_s$ for given $s$ and $\beta$, which proves sufficient for subsequent investigations.

To verify the renormalization operation leaving the long-time dynamics intact, we have compared the energy auto-correlation function [36] both before and after renormalization. Consider an original energy sequence $\{E(t)\}$ with duration $\mathcal{T}$, whose correlation function reads

$$G(t) = \langle (E(0) - \langle E \rangle)(E(t) - \langle E \rangle) \rangle = \langle E(0)E(t) \rangle - \langle E \rangle^2, \tag{6}$$

where $\langle E \rangle = \frac{1}{\mathcal{T}} \sum_1^{\mathcal{T}} E(t)$. Applying the renormalization procedure to this energy sequence yields a new time series $\{E_s(i)\}$, whose correlation function is

$$G_s(t) = \langle E_s(0) E_s(t) \rangle - \langle E_s \rangle^2, \tag{7}$$

According to Eq. (3), we know

$$\langle E_s \rangle = \frac{s}{\mathcal{T}} \sum_i^{\mathcal{T}/s} \frac{1}{s} \sum_{is+1}^{is+s} E(t) = \langle E \rangle. \tag{8}$$

Apparently, $\langle E_s \rangle$ does not change with the temporal renormalization. By combining Eqs. (3), (7), and (8), one readily obtains

$$G_s(t) = \langle E_s(0)E_s(t)\rangle - \langle E_s\rangle^2 = \langle\left(\frac{1}{s}\sum_1^s E(t)\right)\left(\frac{1}{s}\sum_{is+1}^{is+s} E(t)\right)\rangle - \langle E_s\rangle^2$$

$$= \frac{1}{s^2}\sum_{\tau=(i-1)s}^{(i+1)s}(is - |\tau|)G(is+\tau). \tag{9}$$

It can be seen that $G_s(t)$ is nothing but a smoothed version of $G(t)$ centered at $t = is$. Since the timescales of interest typically exceed the renormalization timescale $s$, the correlation for $t > s$ remains less affected. Consequently, the temporal-renormalized partition function $Z_s$ collects its dominant contributions from the long-time dynamics ($t > s$).

**Results and Discussion**: In supercooled liquids, the unusual properties of the system are often governed by slow relaxation processes. These slow relaxation processes yield signals that are considerably weaker compared to fast relaxation processes. The temporal renormalization method provides us with a means to study thermodynamics across different timescales in supercooled liquids. To confirm the effectiveness of the method, we have performed systematic MD simulations of $Cu_{50}Zr_{50}$ alloy, Kob-Andersen binary Lennard-Jones (KA-BLJ) systems [37], and Al supercooled liquids using the open-source code LAMMPS [38] (details of the simulations can be found in the Supplementary Materials). MD simulations confirm that the temporal renormalization does preserve only the long-time correlations. The inset in Fig. 1 shows the energy auto-correlation function $G(t)$ in $Cu_{50}Zr_{50}$ at $T = 800K$. In the absence of renormalization, $G(t)$ exhibits a minimum around 0.05ps, reflecting the characteristic timescale of kinetic or thermal vibrations. Following several oscillations, $G(t)$ plateaus and then gradually decays, indicating the presence of abundant relaxation processes at this timescale. Fig. 1 presents $G_s(t)$ at renormalization scales $s = 10^0, 10^1$, and $10^2$ps. Since $G_s(t)$ curves nearly overlap with the original $G(t)$, they have been vertically offset by 0.25, 0.50, and $0.75eV^2$, respectively. Notably, as $s$ increases, correlations associated with time intervals shorter than $s$ are progressively eliminated, while long-time correlations remain intact. This aligns with the above theoretical analysis and underscores the central objective of the renormalization method. Although Fig. 1 focuses on the $Cu_{50}Zr_{50}$ supercooled liquid, similar behavior is observed in the other two systems.

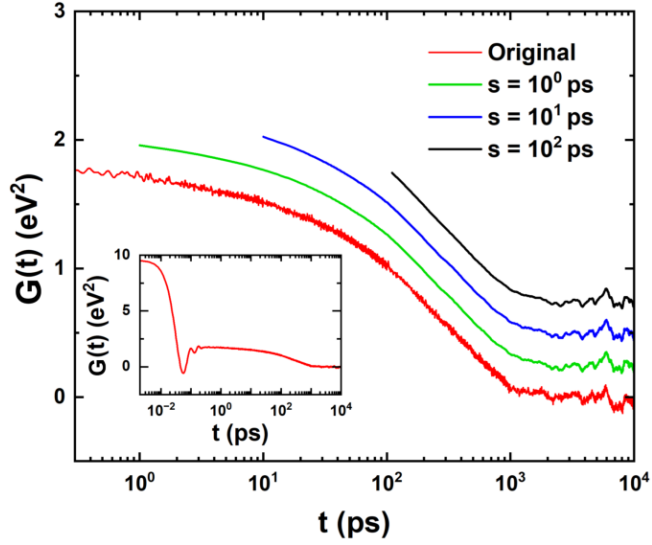

Figure 1. Energy auto-correlation function $G_s(t)$ of $Cu_{50}Zr_{50}$ system at $T = 800K$. As the renormalization timescale $s$ increases, processes below $s$ are erased, yet all correlation behavior of the system at $t > s$ remains preserved. $G_s(t)$ curves at $s = 10^0, 10^1$, and $10^2$ps are shifted upward by $0.25, 0.50$, and $0.75 eV^2$, respectively. The inset displays the complete energy correlation function $G(t)$ without any renormalization applied.

In the supercooled liquid regime, a special renormalization timescale (denoted as the first characteristic renormalization timescale $s_\alpha$) is emerged, in which the second moment of the energy distribution $D_2(s) = \langle E_s^2 \rangle - \langle E_s \rangle^2$ ( more details about $D_2(s)$ can be found in the Supplementary Materials) becomes independent of temperature. Fig. 2 presents contour plots of $D_2(s)$ versus $s$ for three distinct systems. At the supercooled regime, for small $s$, the contour lines tilt downward with decreasing temperature (dashed line), while for large $s$, they shift upward (dotted line). This behavior implies the existence of $s_\alpha$, in which $D_2(s = s_\alpha)$ remains constant across temperatures (solid line). From the contour plots, this special renormalization timescale is determined to be $s_\alpha \approx$ 2.2ps, 3.8ps, and 0.9ps for $Cu_{50}Zr_{50}$, KA-BLJ, and Al, respectively.

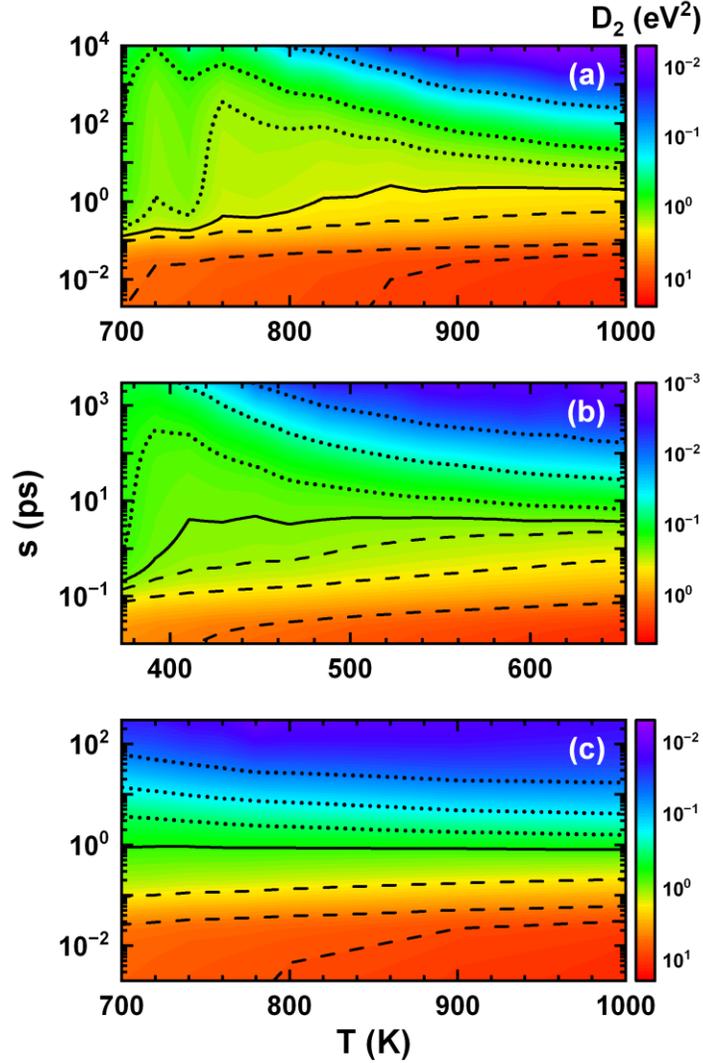

**Figure 2.** Contour plots of $D_2$ versus $s$ and $T$ in (a) $Cu_{50}Zr_{50}$, (b) KA-BLJ, and (c) Al. At the supercooled regime, the contour lines tilt downward with decreasing temperature for small $s$ (dashed line), while for large $s$, the contour lines rise with decreasing temperature (dotted line). Therefore, there exists a special renormalization timescale, i.e., the first characteristic renormalization timescale $s_\alpha$, such that $D_2(s = s_\alpha)$ remains independent of temperature (solid line).

Fig. 3(a) illustrates the temperature dependence of $D_2(s_\alpha)$ in $Cu_{50}Zr_{50}$ at $s_\alpha$. In the high temperature region ($T > 1200K$), where the system exhibits a typical liquid behavior, $D_2(s_\alpha, T)$ decreases monotonically with decreasing temperature. Upon cooling to around 1200 K, the system enters supercooled liquid regime, and $D_2(s_\alpha, T)$ becomes nearly independent of temperature, forming an extended plateau that persists down to approximately 850 K. Below this temperature, the system undergoes a glass transition, and $D_2(s_\alpha, T)$ resumes its decrease with further cooling. Similar phenomenon is also observed in KA-BLJ (Fig. 3(b)) and Al (Fig. 3(c)).

The $D_2(s_\alpha, T)$ plateau emerges exclusively within the supercooled liquid regime, signaling an intrinsic signature of this metastable state. Because the evolution of $D_2$ with the renormalization timescale $s$ encodes the slow dynamics in supercooled liquid, whereas the temperature-insensitive value of $D_2(s_\alpha)$ mirrors the special underlying thermodynamic behavior. The plateau therefore marks a dynamical–thermodynamic connection that is peculiar to the supercooled state.

In conventional renormalization group language, a fixed point denotes a Hamiltonian that is invariant under successive coarse-graining steps [34]. A critical point is the paradigmatic examples. The plateau identified here could be the analogue of such a fixed point under the temporal renormalization. At the special timescale $s_\alpha$, the thermodynamic response functions become temperature-independent, i.e. the system looks "thermodynamically frozen". Supercooled liquids sit precisely on this time-fixed point, implying that the glass transition may be a phase-like change enacted in the temporal domain, rather than spatial one.

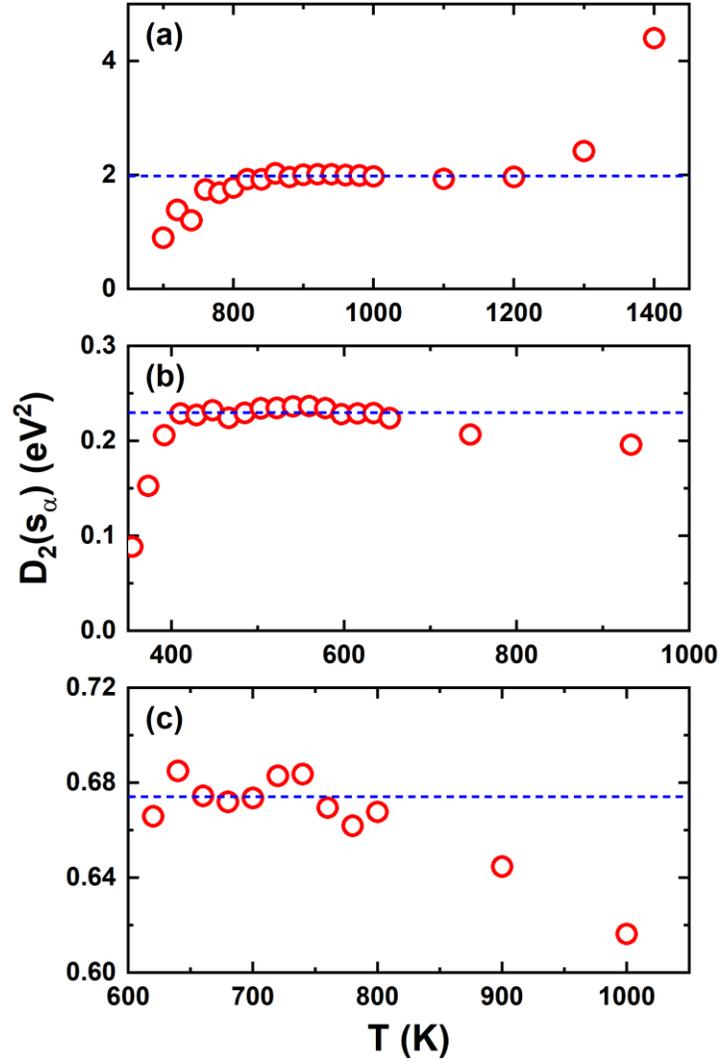

**Figure 3.** The renormalized energy fluctuations $D_2$ obtained at the first characteristic renormalization timescale $s_\alpha$. The three panels show results for (a) $Cu_{50}Zr_{50}$, (b) KA-BLJ, and (c) Al, respectively. In the supercooled liquid region, $D_2(s_\alpha)$ lies at the same height (blue dashed line) and does not vary with temperature.

    This result yields an intriguing picture: at $s_\alpha$, the dynamical behavior in the supercooled liquid contributes identically to the energy fluctuations, implying that the long-time dynamics behavior follows the identical patterns. Furthermore, the specific heat at $s_\alpha$ is given by

$$C(s_\alpha, T) = \frac{D_2(s_\alpha, T)}{k_B T^2} \propto T^{-2}. \tag{10}$$

We further speculate that this simple scaling relation for the renormalized specific heat maybe universally present in supercooled liquids and serve as an intrinsic characteristic of such systems.

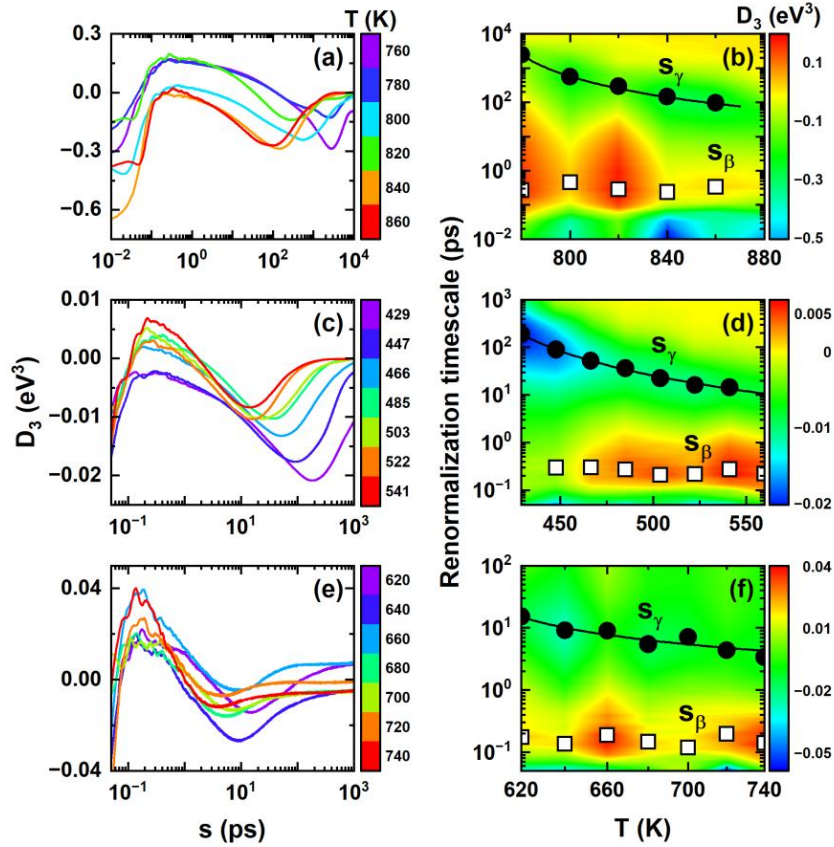

**Figure 4. Variation of $D_3$ with $s$ within the supercooled liquid regime for (a) Cu$_{50}$Zr$_{50}$, (c) KA-BLJ, and (e) Al. As $s$ increases, two characteristic renormalization timescales ($s_\beta$ and $s_\gamma$) successively emerge, at which $D_3$ successively exhibits a maximum and a minimum, respectively. (b), (d), and (f) show the contour plots of $D_3$ for Cu$_{50}$Zr$_{50}$, KA-BLJ, and Al, respectively. Open squares mark the second characteristic renormalization timescale $s_\beta$, which remains nearly independent of temperature; Filled circles denote the third characteristic renormalization timescale $s_\gamma$, which follows a power-law scaling relation. The solid lines in the figure represent fits of $s_\gamma$ using Eq. (11). The fitting critical temperatures**

are $T_c$ = 762K for Cu$_{50}$Zr$_{50}$, 385K for KA-BLJ, and 590K for Al, which are very close to the glass transition temperatures $T_g$.

Fig. 4 illustrates the third moment of the energy distribution $D_3 = \langle (E_s - \langle E_s \rangle)^3 \rangle$ (more details about $D_3(s)$ can be found in the Supplementary Materials) with respect to $s$ at different temperatures for the three systems. At small values of $s$, $D_3$ remains consistently negative, indicating a left-skewed distribution of energies. This suggests that the system predominantly occupies higher-energy states, with a pronounced tail extending into the low energy region. As $s$ increases, $D_3$ gradually rises and reaches a maximum at the second characteristic renormalization timescale (denoted as $s_\beta$, marked by open squares in Fig. 4). Notably, $s_\beta$ appears to be almost independent of temperature. Since $s_\beta$ closely corresponds to the timescale of thermal vibrations, we speculate that at this renormalization timescale, the contribution of vibrational entropy to the free energy may be effectively eliminated. Consequently, the energy distribution approaches a Maxwell-Boltzmann form (which becomes Gaussian in the thermodynamic limit), resulting in $D_3(s_\beta)$ being slightly positive or approximately zero.

Further increasing $s$ reveals the third characteristic renormalization timescale $s_\gamma$, corresponding to the minimum value of $D_3(s = s_\gamma)$ (filled circles in Fig. 4). We find that $s_\gamma$ obeys the power-law scaling relation

$$s_\gamma \sim |T - T_c|^{-\gamma}. \tag{11}$$

Fitting data points yields $T_c$ = 762K and $\gamma$ = 1.93 for Cu$_{50}$Zr$_{50}$ alloy; $T_c$ = 385.25K and $\gamma$ = 2.07 for KA-BLJ; and $T_c$ = 590K and $\gamma$ = 0.77 for Al. In all three systems, $T_c$ is consistently slightly above the glass transition temperature $T_g$. $s_\gamma$ is therefore a previously unidentified characteristic time of supercooled liquids, distinguished from all others by both its magnitude and its critical exponent (see Fig. S1 and the corresponding discussion in Supplement Materials).

We believe that the scaling behavior associated with $s_\alpha$ and $s_\gamma$, identified in this work may be unique to supercooled liquids. In contrast, standard liquids reside in thermodynamically stable states characterized by a single minimum in the free energy landscape. This corresponds to a potential energy distribution that closely approximates a single Gaussian, which becomes the standard Gaussian form in the thermodynamic limit and at infinitely high temperatures. Under such conditions, $D_2(s_\alpha)$ will decrease monotonically with decreasing temperature without exhibiting any plateaus. Similarly, $s_\gamma$ cannot exist in the standard liquids neither. Although standard liquids can also explore multiple metabasins, the characteristic timescale from one metabasin to other are extremely short, comparable to thermal vibration timescales. As a result, in standard liquids, the energy correlation function rapidly decays to zero beyond the characteristic timescale of thermal fluctuations. At this point, the temporal renormalized energy distribution remains very close to a Gaussian distribution. However, for a Gaussian distribution, $D_3$ is always zero, and thus no extreme values exist.

**Summary**: we propose a temporal renormalization scheme. Applied to supercooled liquids, we have identified critical-behavior of supercooled liquid with three characteristic renormalization timescales. 1) A fixed-like point is observed at a specific renormalization scale $s_\alpha$. At this timescale, energy fluctuations remain independent of temperature within the supercooled region. 2) At the second characteristic renormalization timescale $s_\beta$, the third moment of energy distribution $D_3(s_\beta)$ passes through a maximum, and $s_\beta$ is almost independent of temperature. 3) The third characteristic renormalization timescale $s_\gamma$, corresponding to the minimum of $D_3(s_\beta)$, obeys the critical scaling $s_\gamma \sim |T - T_c|^{-\gamma}$, where $T_c$ is slightly higher than but close to the glass transition temperature. The exponent $\gamma$ is distinct from all known static or dynamic exponents, signaling a previously unobserved quasi-criticality. These results reveal a unique "time-domain criticality" in supercooled liquids. The glass transition may be interpreted as a phase-like transition on the temporal domain, and the characteristic renormalization timescales and the associated scaling relations serve as the new characteristic labels of supercooled liquids. This framework provides novel theoretical tools to capture timescale-specific dynamics in glass-forming liquids and, more broadly, in increasingly complex systems.

**Acknowledgements:** This work was supported by the National Key Research and Development Program of China (Grant No. 2022YFA1404603), and by the National Natural Science Foundation of China (Grant No. 12274127 and 12188101).

# References

[1] C. Scalliet, B. Guiselin, and L. Berthier, Thirty milliseconds in the life of a supercooled liquid, Phys. Rev. X **12**, 041028 (2022).

[2] U. Seifert, Stochastic thermodynamics, fluctuation theorems and molecular machines, Rep. Prog. Phys. **75**, 126001 (2012).

[3] S. Emori, U. Bauer, S.-M. Ahn, E. Martinez, and G. S. D. Beach, Current-driven dynamics of chiral ferromagnetic domain walls, Nat. Mater. **12**, 611 (2013).

[4] Y. Tserkovnyak, A. Brataas, and G. E. W. Bauer, Enhanced gilbert damping in thin ferromagnetic films, Phys. Rev. Lett. **88**, 117601 (2002).

[5] E. Beaurepaire, J.-C. Merle, A. Daunois, and J.-Y. Bigot, Ultrafast spin dynamics in ferromagnetic nickel, Phys. Rev. Lett. **76**, 4250 (1996).

[6] G. Grüner, The dynamics of charge-density waves, Rev. Mod. Phys. **60**, 1129 (1988).

[7] M. Strungaru, R. F. L. Evans, and R. W. Chantrell, Substrate contribution to ultrafast spin dynamics in 2D van der Waals magnets, Phys. Rev. Lett. **135**, 076701 (2025).

[8] R. Mondal, L. Rózsa, M. Farle, P. M. Oppeneer, U. Nowak, and M. Cherkasskii, Inertial effects in ultrafast spin dynamics, J. Magn. Magn. Mater. **579**, 170830 (2023).

[9] H. M. Yoo, M. Korkusinski, D. Miravet, K. W. Baldwin, K. West, L. Pfeiffer, P. Hawrylak, and R. C. Ashoori, Time, momentum, and energy resolved pump-probe tunneling spectroscopy of two-dimensional electron systems, Nat. Commun. **14**, 7440 (2023).

[10] L. Caprini, Generalized fluctuation–dissipation relations holding in non-equilibrium dynamics, J. Stat. Mech. Theory Exp. **2021**, 063202 (2021).

[11] W. Gotze and L. Sjogren, Relaxation processes in supercooled liquids, Rep. Prog. Phys. **55**, 241 (1992).

[12] C. A. Angell, Formation of glasses from liquids and biopolymers, Science **267**, 1924 (1995).

[13] P. G. Debenedetti and F. H. Stillinger, Supercooled liquids and the glass transition, Nature **410**, 259 (2001).

[14] Y. Tang, J. Liu, J. Zhang, and P. Zhang, Learning nonequilibrium statistical mechanics and dynamical phase transitions, Nat. Commun. **15**, 1117 (2024).

[15] I. Prigogine, *Non-Equilibrium Statistical Mechanics* (Courier Dover Publications, 2017).

[16] H.-H. Jia et al., Ultrafast relaxation dynamics and nonequilibrium processes in graphite, Phys. Rev. B **111**, 195430 (2025).

[17] C. Dieball and A. ž Godec, Thermodynamic bounds on generalized transport: From single-molecule to bulk observables, Phys. Rev. Lett. **133**, 067101 (2024).

[18] R. K. Pathria, *Statistical Mechanics: International Series of Monographs in Natural Philosophy*, Vol. 45 (Elsevier, 2017).

[19] A. Dechant, J. Garnier-Brun, and S. Sasa, Thermodynamic bounds on correlation times, Phys. Rev. Lett. **131**, 167101 (2023).

[20] J.-C. Delvenne and G. Falasco, Thermokinetic relations, Phys. Rev. E **109**, 014109 (2024).

[21] T. Leadbetter, P. K. Purohit, and C. Reina, A statistical mechanics framework for constructing nonequilibrium thermodynamic models, PNAS Nexus **2**, pgad417 (2023).

[22] T. Matsuo and A. Sonoda, Analysis of entropy production in finitely slow processes between nonequilibrium steady states, Phys. Rev. E **106**, 064119 (2022).

[23] Y. Li, B. C. Abberton, M. Kröger, and W. K. Liu, Challenges in multiscale modeling of polymer dynamics, Polymers **5**, 751 (2013).

[24] A. M. Tsirlin, A. I. Balunov, and I. A. Sukin, Finite-time thermodynamics: Problems, approaches, and results, Entropy Int. Interdiscip. J. Entropy Inf. Stud. **27**, 649 (2025).

[25] M. Schweizer and L. M. C. Sagis, Nonequilibrium thermodynamics of nucleation, J. Chem. Phys. **141**, 224102 (2014).

[26] C. Schütte, S. Klus, and C. Hartmann, Overcoming the timescale barrier in molecular dynamics: Transfer operators, variational principles and machine learning, Acta Numer. **32**, 517 (2023).

[27] V. Lubchenko and P. G. Wolynes, Theory of aging in structural glasses, J. Chem. Phys. **121**, 2852 (2004).

[28] W. H. Wang, Dynamic relaxations and relaxation-property relationships in metallic glasses, Prog. Mater. Sci. **106**, 100561 (2019).

[29] L. Berthier and G. Biroli, Theoretical perspective on the glass transition and amorphous materials, Rev. Mod. Phys. **83**, 587 (2011).

[30] M. Gerry, M. J. Kewming, and D. Segal, Understanding multiple timescales in quantum dissipative dynamics: Insights from quantum trajectories, Phys. Rev. Res. **6**, 033106 (2024).

[31] M. te Vrugt and R. Wittkowski, Mori-Zwanzig projection operator formalism for far-from-equilibrium systems with time-dependent Hamiltonians, Phys. Rev. E **99**, 062118 (2019).

[32] G. Hummer and A. Szabo, Free energy reconstruction from nonequilibrium single-molecule pulling experiments, Proc. Natl. Acad. Sci. **98**, 3658 (2001).

[33]     C. Jarzynski, Nonequilibrium equality for free energy differences, Phys. Rev. Lett. **78**, 2690 (1997).

[34]     S.-K. Ma, *Modern Theory of Critical Phenomena* (Routledge, 2018).

[35]     L. P. Kadanoff, Scaling laws for Ising models near $T_c$, Phys. Phys. Fiz. **2**, 263 (1966).

[36]     J. SCARGLE, Studies in astronomical time-series analysis .3. Fourier-transforms, auto-correlation functions, and cross-correlation functions of unevenly spaced data, Astrophys. J. **343**, 874 (1989).

[37]     W. Kob and H. C. Andersen, Scaling behavior in the β-relaxation regime of a supercooled lennard-jones mixture, Phys. Rev. Lett. **73**, 1376 (1994).

[38]     A. P. Thompson et al., LAMMPS-a flexible simulation tool for particle-based materials modeling at the atomic, meso, and continuum scales, Comput. Phys. Commun. **271**, 108171 (2022).

**Mathematical derivation of Eq. (5) in main text:** To derive Eq. (5), we note that in any realistic system, consecutive microstates are linked by strong dynamical correlations. Introducing the correlation kernel $\eta(s, \beta) \equiv \eta\{E(1), E(2), \cdots, E(s)\}$, which encodes the joint influence of the energies over $s$ consecutive time steps, $P(E_s(i))$ can be written as:

$$P(E_s(i)) \propto \int \eta(s,\beta) \left(\frac{1}{s} \sum_{t=is+1}^{is+s} e^{-\beta E(t)}\right) \delta\left(E_s - \frac{1}{s}\sum_{t=is+1}^{is+s} e^{-\beta E(t)}\right) \prod_{t=is+1}^{is+s} dE(t). \quad (S1)$$

Let $E(t) = E_s(i) + g(t)$ for any $is + 1 \leq t \leq is + s$, then Eq. (S1) transforms into

$$P(E_s(i)) \propto \int \eta(s,\beta) \left(\frac{1}{s} \sum_{t=is+1}^{is+s} e^{-\beta E(t)}\right) \delta\left(\sum_{t=is+1}^{is+s} g(t)\right) \prod_{t=is+1}^{is+s} dE(t) = \Theta(s,\beta) e^{-\beta E_s}, \quad (S2)$$

in which

$$\Theta(s,\beta) = \int \eta(s,\beta) \left(\frac{1}{s} \sum_{t=is+1}^{is+s} e^{-\beta E(t)}\right) \delta\left(\sum_{t=is+1}^{is+s} g(t)\right) \prod_{t=is+1}^{is+s} dg(t). \quad (S3)$$

Although mathematically $g(t)$ may take any real value, $\eta(s, \beta)$ imposes the dynamical constraint, which roughly constrains $g(t)$ in the interval $\sim [-\Delta, +\Delta]$ with $\Delta$ being temperature dependent. Thus, the increment $\Delta g(t) = g(t+1) - g(t)$ is therefore bounded. According to statistical physics, the transition probability between two successive states should be proportional to $e^{-\beta[g(t+1) - g(t)]}$ [1]. And $\Delta$ can be approximated as the fluctuation around the short-time energy average $E_s$, which is proportional to $k_B T$. Let

$$\Delta = \alpha k_B T \quad (S4)$$

where $\alpha$ is a constant at a given temperature. If $g(t)$ is already constrained within $\Delta$, $\eta(s,\beta)$ should vary on a much slower scale than any other factor in Eq. (S1). We may thus replace $\eta(s,\beta)$ by its mean value $\bar{\eta}(s,\beta)$ without altering the underlying physics, whereby Eq. (S3) reduces to

$$\Theta(s,\beta) = \bar{\eta}(s,\beta)\int \frac{1}{s}\prod_{t=2}^{s} e^{\beta g(t)} \prod_{t=2}^{s} \mathrm{d}g(t) + \bar{\eta}(s,\beta)\int \frac{1}{s}\sum_{t=2}^{s} e^{-\beta g(t)} \prod_{t=2}^{s} \mathrm{d}g(t). \quad (S5)$$

Further, Eq. (S5) can be rewritten as:

$$\Theta(s,\beta) \sim \bar{\eta}(s,\beta)\left[\int_{-\Delta}^{\Delta}\frac{1}{s}\prod_{t=2}^{s} e^{\beta g(t)} \prod_{t=2}^{s}\mathrm{d}g(t) + \int_{-\Delta}^{\Delta}\frac{1}{s}\sum_{t=2}^{s} e^{-\beta g(t)} \prod_{t=2}^{s}\mathrm{d}g(t)\right]$$

$$= \bar{\eta}(s,\beta)\left[\frac{1}{s}\left[k_B T\left(e^{\beta\Delta} - e^{-\beta\Delta}\right)\right]^{s-1} - \frac{(s-1)}{s}k_B T(2\Delta)^{s-2}\left(e^{\beta\Delta} - e^{-\beta\Delta}\right)\right]$$

$$= \bar{\eta}(s,\beta)\left[\frac{1}{s}\left(k_B T(e^{\alpha} - e^{-\alpha})\right)^{s-1} - \frac{(s-1)}{s}k_B T(2\alpha k_B T)^{s-2}(e^{\alpha} - e^{-\alpha})\right]. \quad (S6)$$

In the last step, Eq. (S4) is used. Eq. (S6) indicates that at a given temperature, $\Theta(s,\beta)$ is approximately constant and independent of $E_s$. Therefore, the relative magnitudes of $P(E_s(i))$ at the same temperature are independent of $\Theta(s,\beta)$, leading to Eq. (5) of the main text.

**Computational Details :** The $Cu_{50}Zr_{50}$ system contained 2048 atoms, with interatomic interactions described by the potential developed by Mendelev et al. [2]; the KA-BLJ system contained 1000 particles ($N_A : N_B = 800 : 200$) with interactions consistent with the parameter settings in Ref. [3]. For ease of comparison, all KA-BLJ results in this work are converted to real units corresponding to the $Ni_{80}P_{20}$ system [4]. The Al bulk contains 2048 atoms, with interactions modeled by the glue potential proposed by Ercolessi et al. [5]. For all three systems, 10 distinct initial states were prepared at sufficiently high temperatures ($T = 2000$K for $Cu_{50}Zr_{50}$ and Al, and 4500K for KA-BLJ) to ensure the systems reside in a well-defined liquid. Subsequently, the $Cu_{50}Zr_{50}$ and Al liquids were cooled at rates of $10^{10}$ K/s and $10^{12}$ K/s, respectively, under NPT ensemble conditions (external pressure $P = 0$). The KA-BLJ system was cooled at a rate of $1.487 \times 10^9$ K/s under NVT ensemble conditions. In our simulations, the glass transition temperatures for these three systems were 760K for $Cu_{50}Zr_{50}$, 532K for Al, and 363.8K for KA-BLJ, respectively. Subsequently, within the temperature range of interest, the systems transitioned to the NVE ensemble after cooling completion for extended simulations. For $Cu_{50}Zr_{50}$ and Al, each sample underwent 50 ns of NVE simulation, yielding a total of 500ns of data at each temperature of interest. For KA-BLJ, the total sample duration was 300ns. The renormalization calculations in this paper were performed on these 10 samples, and the final results represent the average across all 10 samples.

To demonstrate the power of the temporal renormalization, we take energy fluctuations as an example. After filtering out dynamical processes faster than $s$, the energy fluctuations should accordingly be suppressed. This quantity can be obtained from the second derivative of $\ln Z_s$ with respect to $\beta$, denoted as $D_2(s)$:

$$D_2(s) = \frac{\partial^2 \ln Z_s}{\partial \beta^2} = \langle E_s^2 \rangle - \langle E_s \rangle^2. \tag{S7}$$

$D_2(s)$ is likewise the second moment of the energy distribution. Hence, $D_2(s)$ measures the renormalized energy fluctuation, reflecting thermodynamic behavior at the timescale exceeding $s$. In the extreme case where the renormalization timescale $s$ exceeds all correlation timescales in the system, the energy of every time block becomes equal to $\langle E \rangle$. As a result, the probability distribution $P(E_s)$ collapses to a delta function $\delta(E_s - \langle E \rangle)$, leading to $D_2(s) = 0$.

By applying the temporal renormalization procedure, we can track how the probability distributions of thermodynamic quantities evolve from one timescale to the next. In equilibrium, conventional statistical mechanics predicts that the distribution of thermodynamic quantities is Gaussian. Microscopically, one expects the short-time (high-frequency) modes to dominate the high-energy tail, whereas the long-time (low-frequency) modes control the low-energy sector. As the dynamics on successive temporal slices decouple, the associated distribution is therefore progressively reshaped. To measure the resulting departure from Gaussian form, we calculate the third derivative of $\ln Z_s$ with respect to $\beta$:

$$D_3(s) = -\frac{\partial^3 \ln Z_s}{\partial \beta^3} = \langle (E_s - \langle E_s \rangle)^3 \rangle. \tag{S8}$$

Similar to $D_2(s)$, $D_3(s)$ is precisely the third moment of the distribution of $E_s$ and measures its departure from the Gaussian distribution [6]. For a standard Gaussian distribution, $D_3 = 0$; $D_3 > 0$ signals a right-skewed distribution with an extended high-energy tail, whereas $D_3 < 0$ indicates left skew. Similar to $D_2(s)$, in the limit of large $s$, $D_3(s)$ is expected to vanish.

**The novel characteristic renormalization timescale:** To demonstrate $s_\gamma$ is indeed a novel characteristic time, Fig. S1 compares $s_\gamma$ with the twice energy correlation time ($2\tau_E$), the $\alpha$-relaxation time ($\tau_\alpha$), and the skewness characteristic time ($\Delta_{\text{CTS}}$) [4]. It is evident that $s_\gamma$ differs significantly in magnitude from both $\tau_\alpha$ and $\Delta_{\text{CTS}}$. More importantly, although each quantity can be described by a power law, their exponents differ markedly, confirming that $s_\gamma$ captures a distinct dynamical feature.

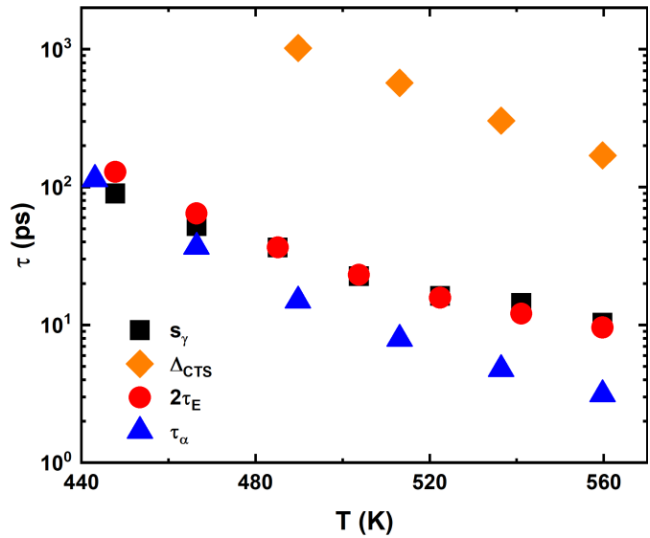

**Figure S1. Comparison of the energy correlation time ($\tau_E$), $\alpha$-relaxation time ($\tau_\alpha$), skewness characteristic time ($\Delta_{\text{CTS}}$), and the characteristic time $s_\gamma$ discovered in this study at different temperatures in KA-BLJ.**

We find that $s_\gamma$ is approximately twice the energy correlation time $\tau_E$, where $\tau_E$ is the characteristic time obtained by fitting the energy correlation function to a stretched exponential equation [7,8]. As shown in Fig. S1, $s_\gamma$ is very close to $2\tau_E$. This may arise from the fact that the variation of the energy distribution with $s$ is inherently determined by the degree of dynamics correlation [4]. As the temperature of the supercooled liquid decreases, the energy correlation time of the system becomes increasingly longer. Once phase-like breaking or dynamic trapping occurs, $s_\gamma$ eventually diverges. That is, at the critical temperature $T_c$, the energy correlation time approaches infinity.

# References

[1]     D. Frenkel and B. Smit, *Understanding Molecular Simulation: From Algorithms to Applications* (Elsevier, 2023).

[2]     M. Mendelev, Y. Sun, F. Zhang, C.-Z. Wang, and K.-M. Ho, Development of a semi-empirical potential suitable for molecular dynamics simulation of vitrification in Cu-Zr alloys, J. Chem. Phys. **151**, (2019).

[3]     W. Kob and H. C. Andersen, Testing mode-coupling theory for a supercooled binary Lennard-Jones mixture I: The van Hove correlation function, Phys. Rev. E **51**, 4626 (1995).

[4]     B. Zhang, D. M. Zhang, D. Y. Sun, and X. G. Gong, Scaling in kinetics of supercooled liquids, Phys. Rev. E **111**, 045417 (2025).

[5]     F. Ercolessi and J. B. Adams, Interatomic potentials from first-principles calculations: the force-matching method, Europhys. Lett. **26**, 583 (1994).

[6]     G. Brys, M. Hubert, and A. Struyf, A robust measure of skewness, J. Comput. Graph. Stat. **13**, 996 (2004).

[7]     J. Phillips, Stretched exponential relaxation in molecular and electronic glasses, Rep. Prog. Phys. **59**, 1133 (1996).

[8]     D. M. Zhang, D. Y. Sun, and X. G. Gong, Mean-squared energy difference for exploring potential energy landscapes of supercooled liquids, Chin. Phys. Lett. **42**, 056301 (2025).